# Human-powered Sorts and Joins

Adam Marcus   Eugene Wu   David Karger   Samuel Madden   Robert Miller
MIT CSAIL
{marcua,sirrice,karger,madden,rcm}@csail.mit.edu

## ABSTRACT

Crowdsourcing markets like Amazon's Mechanical Turk (MTurk) make it possible to task people with small jobs, such as labeling images or looking up phone numbers, via a programmatic interface. MTurk tasks for processing datasets with humans are currently designed with significant reimplementation of common workflows and ad-hoc selection of parameters such as price to pay per task. We describe how we have integrated crowds into a declarative workflow engine called Qurk to reduce the burden on workflow designers. In this paper, we focus on how to use humans to compare items for sorting and joining data, two of the most common operations in DBMSs. We describe our basic query interface and the user interface of the tasks we post to MTurk. We also propose a number of optimizations, including task batching, replacing pairwise comparisons with numerical ratings, and pre-filtering tables before joining them, which dramatically reduce the overall cost of running sorts and joins on the crowd. In an experiment joining two sets of images, we reduce the overall cost from $67 in a naive implementation to about $3, without substantially affecting accuracy or latency. In an end-to-end experiment, we reduced cost by a factor of 14.5.

## 1. INTRODUCTION

Crowd-sourced marketplaces such as Amazon's Mechanical Turk make it possible to recruit large numbers of people to complete small tasks that are difficult for computers to do, such as transcribing an audio snippet or finding a person's phone number on the Internet. Employers submit jobs (Human Intelligence Tasks, or HITs in MTurk parlance) as HTML forms requesting some information or input from workers. Workers (called Turkers on MTurk) perform the tasks, input their answers, and receive a small payment (specified by the employer) in return (typically 1–5 cents).

These marketplaces are increasingly widely used. Crowdflower, a startup company that builds tools to help companies use MTurk and other crowdsourcing platforms now claims to more than 1 million tasks per day to more than 1 million workers and has raised $17M+ in venture capital. CastingWords, a transcription service, uses MTurk to automate audio transcription tasks. Novel academic projects include a word processor with crowdsourced editors [1] and a mobile phone application that enables crowd workers to identify items in images taken by blind users [2].

There are several reasons that systems like MTurk are of interest to database researchers. First, MTurk developers often implement tasks that involve familiar database operations such as filtering, sorting, and joining datasets. For example, it is common for MTurk workflows to filter datasets to find images or audio on a specific subject, or rank such data based on workers' subjective opinion. Programmers currently waste considerable effort re-implementing these operations because reusable implementations do not exist. Furthermore, existing database implementations of these operators cannot be reused, because they are not designed to execute and optimize over crowd workers.

A second opportunity for database researchers is in query optimization. Human workers periodically introduce mistakes, require compensation or incentives, and take longer than traditional silicon-based operators. Currently, workflow designers perform ad-hoc parameter tuning when deciding how many assignments of each HIT to post in order to increase answer confidence, how much to pay per task, and how to combine several human-powered operators (e.g., multiple filters) together into one HIT. These parameters are amenable to cost-based optimization, and introduce an exciting new landscape for query optimization and execution research.

To address these opportunities, we have built *Qurk* [11], a declarative query processing system designed to run queries over a crowd of workers, with crowd-based filter, join, and sort operators that optimize for some of the parameters described above. Qurk's executor can choose the best implementation or user interface for different operators depending on the type of question or properties of the data. The executor combines human computation and traditional relational processing (e.g., filtering images by date before presenting them to the crowd). Qurk's declarative interface enables platform independence with respect to the crowd providing work. Finally, Qurk automatically translates queries into HITs and collects the answers in tabular form as they are completed by workers.

Several other groups, including Berkeley [5] and Stanford [13] have also proposed crowd-oriented database systems motivated by the advantages of a declarative approach. These initial proposals, including our own [11], presented basic system architectures and data models, and described some of the challenges of building such a crowd-sourced database. The proposals, however, did not explore the variety of possible implementations of relational operators as tasks on a crowd such as MTurk.

In this paper, we focus on the implementation of two of the most important database operators, joins and sorts, in Qurk. We believe we are the first to systematically study the implementation of these operators in a crowdsourced database. The human-powered versions of these operators are important because they appear everywhere. For example, information integration and deduplication can be stated as a join between two datasets, one with canonical identifiers for entities, and the other with alternate identifiers. Human-powered sorts are widespread as well. Each time a user provides a rating, product review, or votes on a user-generated content website, they are contributing to a human-powered ORDER BY.

Sorts and joins are challenging to implement because there are a variety of ways they can be implemented as HITs. For example,

to order a list of images, we might ask users to compare groups of images. Alternatively, we might ask users for numerical ratings for each image. We would then use the comparisons or scores to compute the order. The interfaces for sorting are quite different, require a different number of total HITs and result in different answers. Similarly, we explore a variety of ways to issue HITs that compare objects for computing a join, and study answer quality generated by different interfaces on a range of datasets.

Besides describing these implementation alternatives, we also explore optimizations to compute a result in a smaller number of HITs, which reduces query cost and sometimes latency. Specifically, we look at:

- *Batching:* We can issue HITs that ask users to process a variable number of records. Larger batches reduce the number of HITs, but may negatively impact answer quality or latency.
- *Worker agreement:* Workers make mistakes, disagree, and attempt to game the marketplace by doing a minimal amount of work. We evaluate several metrics to compute answer and worker quality, and inter-worker agreement.
- *Join pre-filtering:* There are often preconditions that must be true before two items can be joined. For example, two people are different if they have different genders. We introduce a way for users to specify such filters, which require a linear pass by the crowd over each table being joined, but allow us to avoid a full cross-product when computing the join.
- *Hybrid sort:* When sorting, asking users to rate items requires fewer tasks than directly comparing pairs of objects, but produces a less thorough ordering. We introduce a hybrid algorithm that uses rating to roughly order items, and iteratively improves that ordering by using comparisons to improve the order of objects with similar ratings.

Our join optimizations result in more than an order-of-magnitude (from $67 to $3 on a join of photos) cost reduction while maintaining result accuracy and latency. For sorts, we show that ranking (which requires a number of HITs linear in the input size) costs dramatically less than ordering (which requires a number of HITs quadratic in the input size), and produces comparable results in many cases. Finally, in an end-to-end test, we show that our optimizations can reduce by a factor of 14 the number of HITs required to join images of actors and rank-order them.

In addition to describing these specific operators and optimizations, we review the design of Qurk (originally described in our CIDR paper [11]) and present several extensions to our basic system model that we have developed as we implemented the system.

## 2. LANGUAGE OVERVIEW AND SYSTEM

In this section, we describe the query language and implementation of Qurk. An initial version of this design appeared in a short paper [11], though the design has since been refined.

### 2.1 Data Model and Query Language

This section describes the Qurk data model and query language, and focuses on how joins and sorts are expressed through a series of queries. Our examples have workers provide us with information about various images. We use image examples for consistency of explanation, and because databases typically do not perform processing over images. Qurk's use cases are not limited to processing images. Franklin et al. [5] show how human computation-aware joins can be used for entity disambiguation, and we explore using workers to rate a video in Section 5.

The basic data model is relational, with user-defined scalar and table functions (UDFs) used to retrieve data from the crowd. Rather than requiring users to implement these UDFs in terms of raw HTML forms or low-level declarative code, most UDFs are implemented using one of several pre-defined `Task` templates that specify information about how Qurk should present questions to the crowd.

To illustrate a simple example, suppose we have a table of celebrities, with schema `celeb(name text, img url)`.

We want the crowd to filter this table and find celebrities that are female. We would write:

```
SELECT c.name
FROM celeb AS c
WHERE isFemale(c)
```

With `isFemale` defined as follows:

```
TASK isFemale(field) TYPE Filter:
   Prompt: "<table><tr> \
      <td></td> \
      <td>Is the person in the image a woman?</td> \
    </tr></table>", tuple[field]
   YesText: "Yes"
   NoText: "No"
   Combiner: MajorityVote
```

Tasks have types, which define the low-level implementation and interface to be generated. Filter tasks take tuples as input, and produce tuples that users indicate satisfy the question specified in the `Prompt` field. Here, `Prompt` is simply an HTML block into which the programmer can substitute fields from the tuple being filtered. This tuple is available via the `tuple` variable; its fields are accessed via the use of field names in square brackets. In this case, the question shows an image of the celebrity and asks if they are female. The `YesText` and `NoText` fields allow developers to specify the titles of buttons to answer the question.

Since workers make mistakes, generate unusual answers, or attempt to game the system by performing tasks without thinking to get paid quickly, it is valuable to ask multiple workers for answers. We allow users to specify how many responses are desired. By default we send jobs to 5 workers. Users can specify if they want more workers to answer, and in our experiments we measure the effect of this value on answer quality. We also explore algorithms for adaptively deciding whether another answer is needed in Section 6.

The `Combiner` field specifies a function that determines how to combine multiple responses into one answer. In addition to providing a `MajorityVote` combiner, which returns the most popular answer, we have implemented the method described by Ipeirotis et al. [6]. This method, which we call `QualityAdjust`, identifies spammers and worker bias, and iteratively adjusts answer confidence accordingly in an Expectation Maximization-like fashion.

Filters describe how to ask a worker about one tuple. The query compiler and optimizer can choose to repeat the `Prompts` for several tuples at once. This allows workers to perform several filter operations on records from the same table in a single HIT.

Advanced users of Qurk can define their own tasks that, for example, generate specialized UIs. However, these custom UIs require additional effort if one wishes to take advantage of optimizations such as batching.

### 2.2 Generative Tasks

Filter tasks have a constrained user interface for providing a response. Often, a task requires workers to generate unconstrained input, such as producing a label for an image or finding a phone number. In these situations, we must normalize worker responses to better take advantage of multiple worker responses. Since generative tasks can have workers generate data for multiple fields and return tuples, this is a way to generate tables of data.

For example, say we have a table of animal photos: `animals(id integer, img url)`. We wish to ask workers to provide us with the common name and species of each animal:

```
SELECT id, animalInfo(img).common,
           animalInfo(img).species
FROM animals AS a
```

In this case, `animalInfo` is a generative UDF which returns two fields, `common` with the common name, and `species` with the species name.

```
TASK animalInfo(field) TYPE Generative:
    Prompt: "<table><tr> \
               <td> \
               <td>What is the common name \
                   and species of this animal? \
             </table>", tuple[field]
    Fields: {
       common: { Response: Text("Common name")
                 Combiner: MajorityVote,
                 Normalizer: LowercaseSingleSpace },
       species: { Response: Text("Species"),
                  Combiner: MajorityVote,
                  Normalizer: LowercaseSingleSpace }
    }
```

A generative task provides a `Prompt` for asking a question, much like a filter task. It can return a tuple with fields specified in the `Fields` parameter. Just like the filter task, we can combine the work with a `Combiner`. We also introduce a `Normalizer`, which takes the text input from workers and normalizes it by lowercasing and single-spacing it, which makes the combiner more effective at aggregating responses.

### 2.3 Sorts

Sorts are implemented through UDFs specified in the ORDER BY clause. Suppose, for example, we have a table of images of squares of different sizes, with schema `squares(label text, img url)`. To order these by the area of the square, we write:

```
SELECT squares.label
FROM squares
ORDER BY squareSorter(img)
```

where the task definition for `squareSorter` is as follows.

```
TASK squareSorter(field) TYPE Rank:
    SingularName: "square"
    PluralName: "squares"
    OrderDimensionName: "area"
    LeastName: "smallest"
    MostName: "largest"
    Html: "",tuple[field]
```

As we discuss in Section 4, Qurk uses one of several different interfaces for ordering elements. One version asks users to order small subsets of elements; the other version asks users to provide a numerical ranking for each element. The Rank task asks the developer to specify a set of labels that are used to populate these different interfaces. In the case of comparing several squares, the above text will generate an interface like the one shown in Figure 5.

As with filters, tasks like Rank specified in the ORDER BY clause can ask users to provide ordering information about several records from the input relation in a single HIT. This allows our interface to batch together several tuples for a worker to process.

A Rank UDF can also be used to implement top-K (via a LIMIT clause) and MAX/MIN aggregates. For top-K, we simply perform a complete sort and extract the top K items. For MAX/MIN, we use an interface that extracts the best element from a batch at a time.

### 2.4 Joins and Feature Extraction

The basic implementation of joins is similar to that for sorts and filters. Suppose we want to join a table of images with schema `photos(img url)` with the celebrities table defined above:

```
SELECT c.name
FROM celeb c JOIN photos p
ON samePerson(c.img,p.img)
```

The `samePerson` predicate is an equijoin task, as follows:

```
TASK samePerson(f1, f2) TYPE EquiJoin:
    SingluarName: "celebrity"
    PluralName: "celebrities"
    LeftPreview: "",tuple1[f1]
    LeftNormal: "",tuple1[f1]
    RightPreview: "",tuple2[f2]
    RightNormal: "",tuple2[f2]
    Combiner: MajorityVote
```

The fields in this task are used to generate one of several different join interfaces that is presented to the user. The basic idea with these interfaces is to ask users to compare pairs of elements from the two tables (accessed through the `tuple1` and `tuple2` variables); these pairs are used to generate join results. As with sorting and filter, Qurk can automatically batch together several join tasks into one HIT. A sample interface is shown in Figure 2a.

As we discuss in Section 3.2, we often wish to extract features of items being joined together to filter potential join candidates down, and allow us to avoid computing a cross product. Some features may not be useful for accurately trimming the cross product, and so we introduce a syntax for users to suggest features for filtering that may or may not be used (as we discuss in Section 3.2, the system automatically selects which features to apply.)

We supplement traditional join syntax with a `POSSIBLY` keyword that indicates the features that may help filter the join. For example, the query:

```
SELECT c.name
FROM celeb c JOIN photos p
ON samePerson(c.img,p.img)
AND POSSIBLY gender(c.img) = gender(p.img)
AND POSSIBLY hairColor(c.img) = hairColor(p.img)
AND POSSIBLY skinColor(c.img) = skinColor(p.img)
```

joins the `celeb` and `photos` table as above. The additional `POSSIBLY` clause filters extract gender, hair color, and skin color from images being joined and are used to reduce the number of join candidates that the join considers. Specifically, the system only asks users to join elements from the two tables if all of the predicates in the `POSSIBLY` clause it tries are satisfied (it may not try all predicates.) These predicates can be applied in a linear scan of the tables, avoiding a cross product that might otherwise result. Here, `gender`, `hairColor`, and `skinColor` are UDFs that return one of several possible values (rather than table functions as with the previous UDFs). For example:

```
TASK gender(field) TYPE Generative:
    Prompt: "<table><tr> \
            <td> \
            <td>What is this person's gender? \
          </table>", tuple[field]
    Response: Radio("Gender",
               ["Male","Female",UNKNOWN])
    Combiner: MajorityVote
```

In contrast to the `animalInfo` generative task, note that this generative task only has one field, so it omits the `Fields` parameter. Additionally, the field does not require a `Normalizer` because it has a constrained input space.

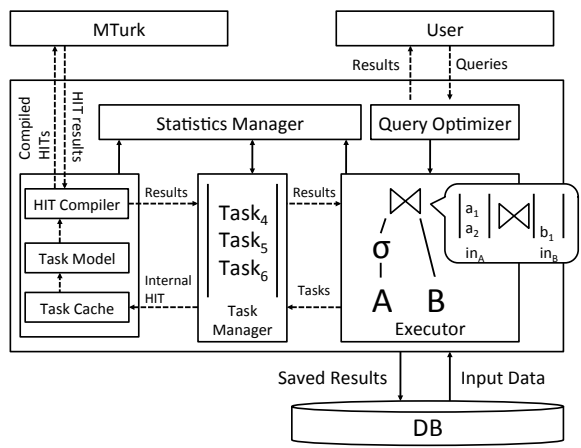

Figure 1: The Qurk system architecture.

It is possible for feature extraction interfaces to generate a special value UNKNOWN, which indicates a worker could not tell its value. This special value is equal to any other value, so that an UNKNOWN value does not remove potential join candidates.

### 2.5 HIT Generation

The above queries are translated into HITs that are issued to the underlying crowd. It's important to generate tasks in a way that keeps the total number of HITs generated down. For example, as in a traditional database, it's better to filter tables before joining them. Query planning in Qurk is done in a way similar to conventional logical to physical query plan generation; a query is translated into a plan-tree that processes input tables in a bottom-up fashion. Relational operations that can be performed by a computer rather than humans are pushed down the query plan as far as possible (including pushing non-HIT joins below HIT-based filters when possible.)

The system generates HITs for all non-join WHERE clause expressions first, and then as those expressions produce results, feeds them into join operators, which in turn produce HITs that are fed to successive operators. As with traditional query execution, HITs for conjuncts (ANDs) of filters are issued serially, while disjuncts (ORs) are issued in parallel. After filters, joins are executed left-deep, with results of lower joins being pipelined into higher joins. Qurk currently lacks selectivity estimation, so it orders filters and joins as they appear in the query.

### 2.6 Architecture and Implementation

In this section, we briefly describe the architecture of Qurk and provide a few details about its implementation.

The basic architecture is shown in Figure 1. Queries are issued through the Query Optimizer, which generates a logical plan and begins executing it in the Executor. The executor runs the plan, generating tasks according to the rules in Section 2.5. These tasks are sent to the Task Manager, which applies batching and other optimizations and dispatches them to the Task Cache/Model/HIT Compiler, which first checks to see if the HIT is cached and if not generates HTML for the HIT and dispatches it to the crowd. As answers come back, they are cached, extracted from their HTML forms, and fed to the executor, which sends the results to the operators that operate on them (or to the user). These operators in turn generate more tasks.

In our implementation, each operator runs in its own thread, asynchronously consuming results from input queues and sending tasks to the Task Manager. Qurk is implemented as a Scala workflow engine with several input types including relational databases and tab-delimited text files. We created several interface prototypes and experiments in Python using the Django web framework.

**Pricing Tasks:** Our current Qurk implementation runs on top of Mechanical Turk. We pay a fixed value per HIT ($0.01 in our experiments). Research by Mason and Watts has suggested that workers on Mechanical Turk do not do particularly higher quality work for higher priced tasks [12]. Mason and Watts also find that workers increase the amount of work they perform with an increase in wage, suggesting that Turkers have an internal model of how much money their work is worth. In all of our experiments, the basic tasks we perform are quick enough that users will do several of them for $0.01, which means we can batch several tasks into a single HIT. Paying more per HIT would allow us to perform more batching, but the degree of additional batching would scale linearly with the additional money we pay, which wouldn't save us money.

**Objective Function:** Because we pay a fixed value per HIT, our system currently uses a simple objective function: minimize the total number of HITs required to fully process a query subject to the constraint that query answers are actually produced[1]. The constraint arises because certain optimizations we apply, like batching, will eventually lead to HITs that are too time-consuming for users to be willing to do for $0.01.

**Batching:** Our system automatically applies two types of batching to tasks: *merging*, where we generate a single HIT that applies a given task (operator) to multiple tuples, and *combining*, where we generate a single HIT that applies several tasks (generally only filters and generative tasks) to the same tuple. Both of these optimizations have the effect of reducing the total number of HITs[2]. We discuss our approach to batching sorts and joins in more detail in the next two sections; for filters and generative tasks, batches are generated by concatenating the HTML forms for multiple tasks together onto the single web page presented to the user.

**HIT Groups:** In addition to batching several tasks into a single HIT, our system groups together (batched) HITs from the same operator into groups that are sent to Mechanical Turk as a single HIT group. This is done because Turkers tend to gravitate toward HIT groups with more tasks available in them, as they can more quickly perform work once they are familiar with the interface. In CrowdDB [5], the authors show the effect of HIT group size on task completion rate.

Now that we've presented Qurk's architecture, we describe the implementations and optimizations we developed for joins and sorts.

## 3. JOIN OPERATOR

This section describes several implementations of the join operator, and the details of our feature filtering approach for reducing join complexity. We present a series of experiments to show the quality and performance of different join approaches.

### 3.1 Implementation

The join HIT interface asks a worker to compare elements from two joined relations. Qurk implements a block nested loop join, and uses the results of the HIT comparisons to evaluate whether two elements satisfy the join condition. We do not implement more efficient join algorithms (e.g., hash join or sort-merge join) because we do not have a way to compute item (e.g., picture) hashes for hash joins or item order for sort-merge joins.

---

[1]Other objective functions include maximizing answer quality or minimizing latency. Unfortunately, answer quality is hard to define (the correct answer to many human computation tasks cannot be known). Latency is highly variable, and probably better optimized through low-level optimizations like those used in quikTurkit [2].

[2]For sequences of conjunctive predicates, combining actually does more "work" on people than not combining, since tuples that may have been discarded by the first filter are run through the second filter as well. Still, as long as the first filter does not have 0 selectivity, this will reduce the total number of HITs that have to be run.

The following screenshots and descriptions center around evaluating join predicates on images, but are not limited to image data types. The implementations generalize to any field type that can be displayed in HTML. In this section, we assume the two tables being joined are $R$ and $S$, with cardinalities $|R|$ and $|S|$, respectively.

### 3.1.1 Simple Join

Figure 2a shows an example of a simple join predicate interface called SimpleJoin. In this interface, a single pair of items to be joined is displayed in each HIT along with the join predicate question, and two buttons (*Yes, No*) for whether the predicate evaluates to true or false. This simplest form of a join between tables *R* and *S* requires $|R||S|$ HITs to be evaluated.

### 3.1.2 Naive Batching

Figure 2b shows the simplest form of join batching, called Naive-Batch. In NaiveBatch, we display several pairs vertically. *Yes, No* radio buttons are shown with each pair that is displayed. A *Submit* button at the bottom of the interface allows the worker to submit all of the pairs evaluated in the HIT. If the worker clicks *Submit* without having selected one of *Yes* or *No* for each pair, they are asked to select an option for each unselected pair.

For a batch size of $b$, where $b$ pairs are displayed in each HIT, we can reduce the number of HITs to $\frac{|R||S|}{b}$.

### 3.1.3 Smart Batching

Figure 2c shows a more complex join batching interface called SmartBatch. Two columns of images are displayed, and workers are asked to click on pairs of images that match the join predicate. The first column contains images from table *R* and the second contains images from table *S*.

Once a worker selects a pair, it is displayed in a list to the right of the columns, and can be removed (if added by mistake) by clicking on the pair. All selected pairs are connected by a line. If none of the images match the join predicate, the worker is asked to click a checkbox indicating no matches. In order to submit the HIT, the box must be checked or at least one pair must be selected.

To conserve vertical space, images are not displayed at full size. If a user hovers over an image, it is displayed at the size used in SimpleJoin and NaiveJoin (e.g., in Figure 2c, the mouse is hovering over Notorious B.I.G, who is displayed at full size).

For $r$ images in the first column and $s$ in the second column, we must evaluate $\frac{|R||S|}{rs}$ HITs.

### 3.1.4 Alternative Join Algorithms

There are a number of alternative join algorithms we have not discussed. For example, we could ask workers to label each tuple with a unique identifier of the entity that it represents, and perform a traditional join on the identifier. Our goal is to understand the accuracy-cost tradeoffs of batching and combining, so these alternatives are outside this paper's scope. Our results can still be used to benefit other join algorithms, and we use the idea of labeling tuples for the feature filtering optimization described in Section 3.2.

## 3.2 Feature Filtering Optimization

In Section 2.1, we introduced the POSSIBLY clause to joins for identifying feature-based filters that may reduce the size of a join cross product. This clause allows the developer to specify that some features must be true for the join predicate to evaluate to true. For example, two profile images shouldn't join unless they have the same gender, hair color, and skin color. These predicates allow us to only consider join pairs which match the extracted features.

We now explain the benefit of this filtering. To simplify our analysis, we assume that all filter features are uncorrelated, and that the filters do not emit the value UNKNOWN.

Suppose there are $N$ POSSIBLY clauses added to a join. Let $F = \{F_1, ..., F_N\}$, where $F_i$ is a set that contains the possible values for the feature being compared in POSSIBLY clause $i$. For example, if the $i$th feature is *hairColor*, $F_i = \{$black, brown, blond, white$\}$. Let the probability that feature $i$ (e.g., hair color) has value $j$ (e.g., brown) in table $X$ to be $\rho_{Xij}$. Then, for two tables, $R$ and $S$, the probability that those two tables match on feature $i$ is:

$$\sigma_i = \sum_{j \in F_i} \rho_{Sij} \times \rho_{Rij}$$

In other words, $\sigma_i$ is the *selectivity* of feature $i$. Thus, the selectivity of all expressions in the POSSIBLY clause (assuming the features are independent) is:

$$Sel = \prod_{i \in [1...N]} \sigma_i$$

Feature filtering causes the total number of join HITs that are executed to be a fraction $Sel$ of what would be executed by a join algorithm alone. This benefit comes at the cost of running one linear pass over each table for each feature filter. Of course, the HITs in the linear pass can be batched through merging and combining.

In general, feature filtering is helpful, but there are three possible cases where we may not want to apply a filter: 1) if the additional cost of applying the filter does not justify the reduction in selectivity it offers (e.g., if all of the people in two join tables of images have brown hair); 2) if the feature doesn't actually guarantee that two entities will not join (e.g., because a person has different hair color in two different images); or 3) if the feature is ambiguous (i.e., workers do not agree on its value).

To detect 1), we run the feature filter on a small sample of the data set and estimate selectivity, discarding filters that are not effective. To evaluate 2) for a feature $f$, we also use a sample of both tables, computing the join result $j_{f-}$ with all feature filters except $f$, as well as the join result with $f$, $j_{f+}$. We then measure the fraction $\frac{|j_{f-} - j_{f+}|}{|j_{f-}|}$ and if it is below some threshold, we discard that feature filter clause from the join.

For case 3) (feature ambiguity), we use a measure called inter-rater reliability (IRR), which measures the extent to which workers agree. As a quantitative measure of IRR, we utilize Fleiss' $\kappa$ [4]. Fleiss' $\kappa$ is used for measuring agreement between two or more raters on labeling a set of records with categorical labels (e.g., true or false). It is a number between -1 and 1, where a higher number indicates greater agreement. A $\kappa$ of 0 roughly means that the ratings are what would be expected if the ratings had been sampled randomly from a weighted distribution, where the weights for a category are proportional to the frequency of that category across all records. For feature filters, if we measure $\kappa$ to be below some small positive threshold for a given filter, we discard it from our filter set. Due to our use of Fleiss' $\kappa$, Qurk currently only supports detecting ambiguity for categorical features, although in some cases, range-valued features may be binned into categories.

## 3.3 Experiments

We now explore the various join implementations and the effects of batching and feature filtering. We also explore the quality of worker output as they perform more tasks.

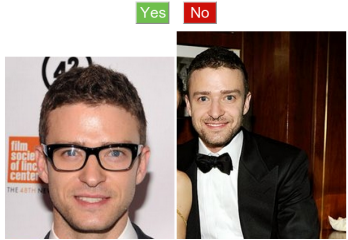
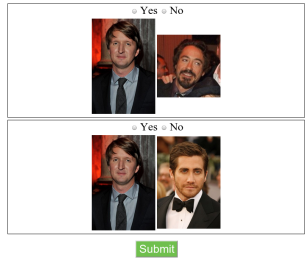
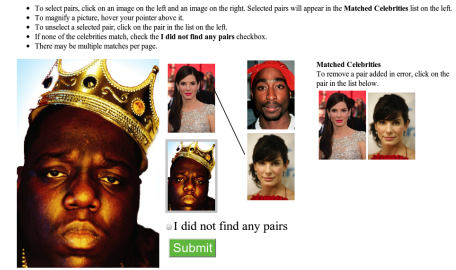

| (a) Simple Join | (b) Naive Batching | (c) Smart Batching |

Figure 2: Three interfaces for the join operator. Higher-resolution screenshots available at http://db.csail.mit.edu/qurk.

| Implementation | True Pos. (MV) | True Pos. (QA) | True Neg (MV) | True Neg (QA) |
|---|---|---|---|---|
| IDEAL | 20 | 20 | 380 | 380 |
| Simple | 19 | 20 | 379 | 376 |
| Naive | 19 | 19 | 380 | 379 |
| Smart | 20 | 20 | 380 | 379 |

Table 1: Baseline comparison of three join algorithms with no batching enabled. Each join matches 20 celebrities in two tables, resulting in 20 image matches (1 per celebrity) and 380 pairs with non-matching celebrities. Results reported for ten assignments aggregated from two trials of five assignments each. With no batching enabled, the algorithms have comparable accuracy.

### 3.3.1 Dataset

In order to test join implementations and feature filtering, we created a *celebrity join dataset*. This dataset contains two tables. The first is celeb(name text, img url), a table of known celebrities, each with a profile photo from IMDB[3]. The second table is photos(id int, img url), with of images of celebrities collected from People Magazine's collection of photos from the 2011 Oscar awards.

Each table contains one image of each celebrity, so joining $N$ corresponding rows from each table naively takes $N^2$ comparisons, and has selectivity $\frac{1}{N}$.

### 3.3.2 Join Implementations

In this section, we study the accuracy, price, and latency of the celebrity join query described in Section 2.4.

We run each of the join implementations twice (Trial #1 and #2) with five assignments for each comparison. This results in ten comparisons per pair. For each pair of trials, We ran one trial in the morning before 11 AM EST, and one in the evening after 7 PM EST, to measure variance in latency at different times of day. All assignments are priced at $0.01, which costs $0.015 per assignment due to Amazon's half-cent HIT commission.

We use the two methods described in Section 2.1 to combine the join responses from each assignment. For MajorityVote, we identify a join pair if the number of positive votes outweighs the negative votes. For QualityAdjust, we generate a corpus that contains each pair's *Yes, No* votes along with the Amazon-specified Turker ID for each vote. We execute the algorithm in [6] for five iterations on the corpus, and parametrize the algorithm to penalize false negatives twice as heavily as false positives.

**Baseline Join Algorithm Comparison:** We first verify that the three join implementations achieve similar accuracy in unbatched form. Table 1 contains the results of the joins of a sample of 20 celebrities and matching photos. The ideal algorithm results in 20 positive matches and 380 negative matches (pairs which do not join). The true positives and negatives for MajorityVote and QualityAdjust on all ten assignments per pair are reported with the prefixes MV and QA, respectively. From these results, it is evident that all approaches work fairly well, with at most 1 photo which was not correctly matched (missing true positive). We show in the next section that using QA and MV is better than trusting any one worker's result.

[3] http://www.imdb.com

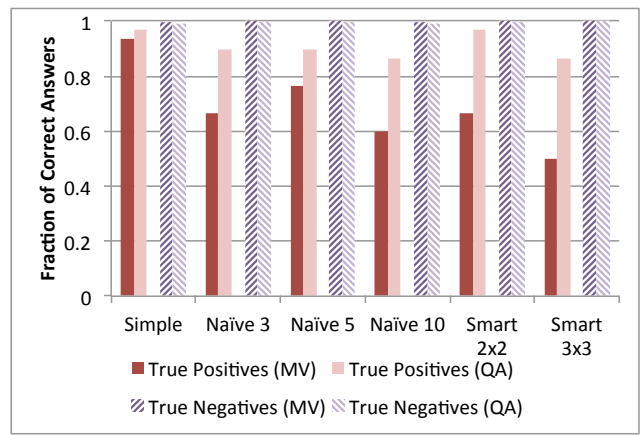

Figure 3: Fraction of correct answers on celebrity join for different batching approaches. Results reported for ten assignments aggregated from two runs of five assignments each. Joins are conducted on two tables with 30 celebrities each, resulting in 30 matches (1 per celebrity) and 870 non-matching join pairs.

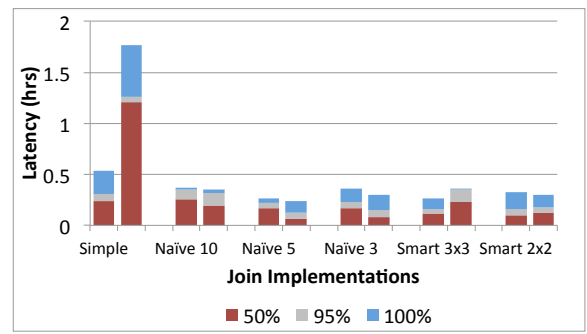

Figure 4: Completion time in hours of the $50^{th}$, $95^{th}$, and $100^{th}$ percentile assignment for variants of celebrity join on two tables with 30 celebrities each.

**Effect of Batching:** In our next experiment, we look at the effect of batching on join quality and price. We compared our simple algorithm to naive batching with 3, 5, and 10 pairs per HIT and smart batching with a $2 \times 2$ and $3 \times 3$ grid, running a celebrity join between two images tables with 30 celebrity photos in each table. The answer quality results are shown in Figure 3. There are several takeaways from this graph.

First, all batching schemes except Smart 2x2, which performs as well as the Simple Join, do have some negative effect on the overall total number of true positives. When using QA, the effect is relatively minor with 1–5 additional false negatives on each of the batching schemes. There is not a significant difference between naive and smart batching. Batching does not significantly affect the overall true negative rate.

Second, QA does better than MV in improving true positive result quality on batched schemes. This is likely because QA includes filters for identifying spammers and sloppy workers, and these larger, batched schemes are more attractive to workers that quickly and inaccurately complete tasks. The overall error rate between two trials of 5 assignments per pair was approximately the same. However, individual trials are more vulnerable to a small number of spammers, which results in higher variance in accuracy.

Third, MV and QA often achieve far higher accuracy as compared to the expected accuracy from asking a single worker for each HIT. In the Simple experiments, the expected true positive rate of an average worker was $235/300 = 78\%$, whereas MV was $93\%$. MV performed the worst in the Smart 3x3 experiments, yet still performed as well the expected true positive rate of $158/300 = 53\%$. In all cases, QA performed significantly better.

We also measured the cost (in dollars) of running the complete join (900 comparisons) for the two trials (with 10 assignments per pair) at a cost of $0.015 per assignment ($0.01 to the worker, $0.005 to Amazon). The cost of a naive join is thus $900 \times \$0.015 \times 10 = \$135.00$. The cost falls proportionally with the degree of batching (e.g., naive 10 reduces cost by a factor of 10, and a 3x3 join reduces cost by a factor of 9), resulting in a cost of around $13.50.

Figure 4 shows end-to-end latency values for the different join implementations, broken down by the time for $50\%$, $95\%$, and $100\%$ percent of the assignments to complete. We observe that a reduction in HITs with batching reduces latency, even though fewer HITs are posted and each HIT contains more work. Both Simple-Join trials were slower than all other runs, but the second Simple-Join trial was particularly slow. This illustrates the difficulty of predicting latency in a system as dynamic as MTurk. Finally, note that in several cases, the last $50\%$ of wait time is spent completing the last $5\%$ of tasks. This occurs because the small number of remaining tasks are less appealing to Turkers looking for long batches of work. Additionally, some Turkers pick up and then abandon tasks, which temporarily blocks other Turkers from starting them.

### 3.3.3 Assignments vs. Accuracy

One concern is that a worker's performance will degrade as they execute more tasks and become bored or less cautious. This is a concern as our results (and those of CrowdDB [5]) suggest that a small number of workers complete a large fraction of tasks.

To test if the amount of work done by a worker is negatively correlated with work quality, we performed a linear regression. For a combination of responses to the two simple $30 \times 30$ join tasks, we fit the number of tasks each worker did with their accuracy ($\frac{\text{correct tasks}}{\text{tasks completed}}$), and found $R^2 = 0.028$, $p < .05$. Accuracy and number of tasks are positively correlated (the slope, $\beta$, is positive), and the correlation explains less than $3\%$ of variance in accuracy. This suggests no strong effect between work done and accuracy.

| Trial # | Combined? | Errors | Saved Comparisons | Join Cost |
|---|---|---|---|---|
| 1 | Y | 1 | 592 | $27.52 |
| 2 | Y | 3 | 623 | $25.05 |
| 1 | N | 5 | 633 | $33.15 |
| 2 | N | 5 | 646 | $32.18 |

Table 2: Feature Filtering Effectiveness.

### 3.3.4 Feature Filtering

Finally, we ran an experiment to measure the effectiveness of feature filtering. In this experiment, we asked workers to choose the hair color, skin color, and gender of each of the 60 images in our two tables. For each feature, we ran two trials with 5 votes per image in each trial, combining answers using majority vote. We also ran two trials with a combined interface where we asked workers to provide all three features at once.

Table 2 shows the effectiveness of applying feature filters in the four trials. We report the number of errors, which is the number of pairs that should have joined (of 30) that didn't pass all three feature filters. We then report the saved comparisons, which is the number of comparisons (of 870) that feature filtering helped avoid. We also report the total join cost with feature filtering. Without feature filters the cost would be $67.50 for 5 assignments per HIT.

From these results, we can see that feature filters substantially reduce the overall cost (by more than a factor of two), and that combining features both reduces cost and error rate. The reason that combining reduces error rate is that in the batched interface, workers were much more likely to get hair color correct than in the non-batched interface. We hypothesize that this is because when asked about all three attributes at once, workers felt that they were doing a simple demographic survey, while when asked solely any one feature (in this case hair color), they may have overanalyzed the task and made more errors.

We now look at the error rate, saved comparisons, and total cost when we omit one feature from the three. The goal of this analysis is to understand whether omitting one of these features might improve join quality by looking at their effectiveness on a small sample of the data as proposed in Section 3.2. The results from this analysis on the first combined trial are shown in Table 3 (all of the trials had the same result). From this table, we can see that omitting features reduces the error rate, and that gender is by far the most effective feature to filter on. From this result, we conclude that hair color should potentially be left out. In fact, hair color was responsible for all of the errors in filtering across all trials.

| Omitted Feature | Errors | Saved Comparisons | Join Cost |
|---|---|---|---|
| Gender | 1 | 356 | $45.30 |
| Hair Color | 0 | 502 | $34.35 |
| Skin Color | 1 | 542 | $31.28 |

Table 3: Leave One Out Analysis for the first combined trial. Removing hair color maintains low cost and avoids false negatives.

To see if we can use inter-rater reliability as a method for determining which attributes are ambiguous, we compute the value of $\kappa$ (as described in Section 3.2) for each of the attributes and trials. The results are shown in Table 4. From the table, we can see that the $\kappa$ value for gender is quite high, indicating the workers generally agree on the gender of photos. The $\kappa$ value for hair is much lower, because many of the celebrities in our photos have dyed hair, and because workers sometimes disagree about blond vs. white hair. Finally, workers agree more about skin color when it is presented in the combined interface, perhaps because they may feel uncomfortable answering questions about skin color in isolation.

Table 4 displays average and standard deviations of $\kappa$ for 50 25% random samples of celebrities in each trial. We see that these $\kappa$ value approximations are near the true $\kappa$ value in each trial, showing that Qurk can use early $\kappa$ values to accurately estimate worker agreement on features without exploring the entire dataset.

From this analysis, we can see that $\kappa$ is a promising metric for automatically determining that hair color (and possibly skin color) should not be used as a feature filter.

| Trial | Sample Size | Combined? | Gender $\kappa$ (std) | Hair $\kappa$ (std) | Skin $\kappa$ (std) |
|---|---|---|---|---|---|
| 1 | 100% | Y | 0.93 | 0.29 | 0.73 |
| 2 | 100% | Y | 0.89 | 0.42 | 0.95 |
| 1 | 100% | N | 0.85 | 0.43 | 0.45 |
| 2 | 100% | N | 0.94 | 0.40 | 0.47 |
| 1 | 25% | Y | 0.93 (0.04) | 0.26 (0.09) | 0.61 (0.37) |
| 2 | 25% | Y | 0.89 (0.06) | 0.40 (0.11) | 0.95 (0.20) |
| 1 | 25% | N | 0.85 (0.07) | 0.45 (0.10) | 0.39 (0.29) |
| 2 | 25% | N | 0.93 (0.06) | 0.38 (0.08) | 0.47 (0.24) |

Table 4: Inter-rater agreement values ($\kappa$) for features. For each trial, we display $\kappa$ calculated on the entire trial's data and on 50 random samples of responses for 25% of celebrities. We report the average and standard deviation for $\kappa$ from the 50 random samples.

### 3.4 Summary

We found that for joins, batching is an effective technique that has small effect on result quality and latency, offering an order-of-magnitude reduction in overall cost. Naive and smart batching perform similarly, with smart 2x2 batching and QA achieving the best accuracy. In Section 5 we show an example of a smart batch run where a 5x5 smart batch interface was acceptable, resulting in a 25x cost reduction. We have never seen such large batches work for naive batching. We found that the QA scheme in [6] significantly improves result quality, particularly when combined with batching, because it effectively filters spammers. Finally, feature filtering offers significant cost savings when a good set of features can be identified. Putting these techniques together, we can see that for celebrity join, feature filtering reduces the join cost from $67.50 to $27.00. Adding batching can further reduce the cost by up to a factor of ten, yielding a final cost for celebrity join of $2.70.

## 4. SORT OPERATOR

Users often want to perform a crowd-powered sort of a dataset, such as "order these variants of a sentence by quality," or "order the images of animals by adult size."

As with joins, the HITs issued by Qurk for sorting do not actually implement the sort algorithm, but provide an algorithm with information it needs by either: 1) comparing pairs of items to each other, or 2) assigning a rating to each item. The Qurk engine then sorts items using pairwise comparisons or their ratings.

### 4.1 Implementations

We now describe our two basic implementations of these ideas, as well as a hybrid algorithm that combines them. We also compare the accuracy and total number of HITs required for each approach.

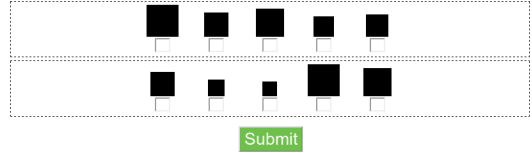

(a) Comparison Sort

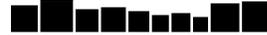

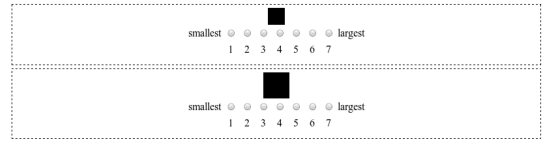

(b) Rating Sort

Figure 5: Two interfaces for the sort operator. Larger screenshots available at http://db.csail.mit.edu/qurk.

#### 4.1.1 Comparison-based

The comparison-based approach (Figure 5a) asks workers to directly specify the ordering of items in the dataset. The naive approach requires $\binom{N}{2}$ tasks per sort assignment, which is expensive for large datasets. While the worst-case number of comparisons is $O(NlogN)$ for traditional sort algorithms, we now explain why we require more comparison tasks.

In practice, because these individual sort HITs are done by different workers, and because tasks themselves may be ambiguous, it can be the case that transitivity of pairwise orderings may not hold. For example, worker 1 may decide that $A > B$ and $B > C$, and worker 2 may decide that $C > A$. One way to resolve such ambiguities is to build a directed graph of items, where there is an edge from item $i$ to item $j$ if $i > j$. We can run a cycle-breaking algorithm on the graph, and perform a topological sort to compute an approximate order. Alternatively, as we do in our implementation, we can compute the number of HITs in which each item was ranked higher than other items. This approach, which we call "head-to-head," provides an intuitively correct ordering on the data, which is identical to the true ordering when there are no cycles.

Cycles also mean that we cannot use algorithms like Quicksort that only perform $O(NlogN)$ comparisons. These algorithms do not compare all elements, and yield unpredictable results in ambiguous situations (which we found while running our experiments).

Instead of comparing a single pair at a time, our interface, shown in Figure 5a, displays groups of $S$ items, and asks the worker to rank items within a group relative to one-another. The result of each task is $\binom{S}{2}$ pairwise comparisons, which reduces the number of HITs to $\frac{N \times (N-1)}{S \times (S-1)}$. Although the number of HITs is large, they can be executed in parallel. We can batch $b$ such groups in a HIT to reduce the number of hits by a factor of $b$.

### 4.1.2 Rating-based

The second approach is to ask workers to rate each item in the dataset along a numerical scale. We then compute the mean of all ratings for each item, and sort the dataset using these means.

Figure 5b illustrates the interface for a single rating task. The worker is shown a single item and asked to rate it along a seven-point Likert Scale [9], which is commonly used for subjective survey data. In order to provide context to assign a rating, we show ten randomly sampled images along the top of the interface. Showing a random selection allows us to give the worker a sense for the dataset without knowing its distribution *a priori*.

The advantage of this approach are that it only requires $O(N)$ HITs. We can batch $b$ ratings in a HIT to reduce the number of HITs by a factor of $b$. The variance of the rating can be reduced by asking more workers to rate the item. The drawback is that each item is rated independently of other items, and the relative ordering of an item pair's mean ratings may not by fully consistent with the ordering that would result if workers directly compared the pair.

### 4.1.3 Hybrid Algorithm

We now propose a hybrid approach that initially orders the data using the rating-based sort and generates a list $L$. Each item $l_i \in L$ has an average rating $\mu_i$, as well as a standard deviation $\sigma_i$ computed from votes used to derive the rating. The idea of our hybrid approach is to iteratively improve $L$ by identifying subsets of $S$ items that may not be accurately ordered and using the comparison-based operator to order them. The user can control the resulting accuracy and cost by specifying the number of iterations (where each iteration requires one additional HIT) to perform.

We explored several techniques for selecting size-$S$ windows for comparisons. We outline three representative approaches:

**Random**: In each iteration, pick $S$ items randomly from $L$.
**Confidence-based**: Let $w_i = \{l_i, ..., l_{i+S}\}$, meaning $w_i$ contains the $S$ consecutive items $l_j \in L$ starting from item $l_i$. For each pair of items $a, b \in w_i$, we have their rating summary statistics $(\mu_a, \sigma_a)$ and $(\mu_b, \sigma_b)$. For $\mu_a < \mu_b$, we compute $\Delta_{a,b}$, the difference between one standard deviation above $\mu_a$ and one standard deviation below $\mu_b$, where $\Delta_{a,b} = \max(\mu_a + \sigma_a - \mu_b - \sigma_b, 0)$. For all windows $w_i$, we then compute $R_i = \sum_{(a,b) \in w_i} \Delta_{a,b}$ and order windows in decreasing order of $R_i$, such that windows with the most standard deviation overlap, and thus least confidence in their ratings, are reordered first.
**Sliding window**: The algorithm picks a window $w_i = \{l_{i \bmod |L|}, ..., l_{(i+S) \bmod |L|}\}$ with $i$ starting at 1. In successive iterations, $i$ is incremented by $t$ (e.g., $i = (i+t)$), which the $\bmod$ operation keeps the range in $[1, |L|]$. If $t$ is not a divisor of $L$, when successive windows wrap around $L$, they will be offset from the previous passes.

## 4.2 Experiments

We now describe experiments that compare the performance and accuracy effects of the Compare and Rate sort implementations, as well as the improvements of our Hybrid optimizations.

The experiments compare the relative similarity of sorted lists using Kendall's Tau ($\tau$), which is a measure used to compute rank-correlation [7]. We use the $\tau - b$ variant, which allows two items to have the same rank order. The value varies between -1 (inverse correlation), 0 (no correlation), and 1 (perfect correlation).

For each pair in Compare, we obtain at least 5 comparisons and take the majority vote of those comparisons. For each item in Rate, we obtain 5 scores, and take the mean of those scores. We ran two trials of each experiment.

### 4.2.1 Datasets

The *squares dataset* contains a synthetically generated set of squares. Each square is $n \times n$ pixels, and the smallest is $20 \times 20$. A dataset of size $N$ contains squares of sizes $\{(20 + 3 * i) \times (20 + 3 * i) | i \in [0, N)\}$. This dataset is designed so that the sort metric (square area) is clearly defined, and we know the correct ordering.

The *animals dataset* contains 25 images of randomly chosen animals ranging from ants to humpback whales. In addition, we added an image of a rock and a dandelion to introduce uncertainty. This is a dataset on which comparisons are less certain, and is designed to show relative accuracies of comparison and rating-based operators.

### 4.2.2 Square Sort Microbenchmarks

In this section, we compare the accuracy, latency, and price for the query described in Section 2.3, in which workers sort squares by their size.

**Comparison batching**. In our first experiment, we sort a dataset with 40 squares by size. We first measure the accuracy of Compare as the group size $S$ varies between 5, 10, and 20. Batches are generated so that every pair of items has at least 5 assignments. Our batch-generation algorithm may generate overlapping groups, so some pairs may be shown more than 5 times. The accuracy is perfect when $S = 5$ and $S = 10$ ($\tau = 1.0$ with respect to a perfect ordering). The rate of workers accepting the tasks dramatically decreases when the group size is above 10 (e.g., the task takes 0.3 hours with group size 5, but more than 1 hour with group size 10.) We stopped the group size 20 experiment after several hours of uncompleted HITs. We discuss this effect in more detail, and ways to deal with it, in Section 6.

**Rating batching**. We then measure the accuracy of the Rate implementation. The interface shows 10 sample squares, sampled randomly from the 40, and varies the batch size from 1 to 10, requiring 40 to 4 HITs, respectively. In all cases, the accuracy is lower than Compare, with an average $\tau$ of 0.78 (strong but not perfect ranking correlation) and standard deviation of 0.058. While increasing batch size to large amounts made HITs less desirable for turkers and eventually increased latency, it did not have a noticeable effect on accuracy. We also found that 5 assignments per HIT resulted in similar accuracy to 10 assignments per HIT, suggesting diminishing returns for this task.

**Rating granularity**. Our next experiment is designed to measure if the granularity of the seven-point Likert scale affects the accuracy of the ordering as the number of distinct items increases. We fix the batch size at 5, and vary the size of the dataset from 20 to 50 in increments of 5. The number of HITs vary from 4 to 10, respectively. As with varying batch size, the dataset size does not significantly impact accuracy (avg $\tau$ 0.798, std 0.042), suggesting that rating granularity is stable with increasing dataset size. While combining 10 assignments from two trials did reduce $\tau$ variance, it did not significantly affect the average.

### 4.2.3 Query Ambiguity: Sort vs. Rank

The square sort microbenchmarks indicate that Compare is more accurate than Rate. In our next experiment, we compare how increasing the ambiguity of sorting tasks affects the accuracy of Rate relative to Compare. The goal is to test the utility of metrics that help predict 1) if the sort task is feasible at all, and 2) how closely Rate corresponds to Compare. The metric we use to answer 1) is a modified version of Fleiss' $\kappa$ (which we used for inter-reliability rating in joins)[4], and the metric to answer 2) is $\tau$. The experiment uses both the *squares* and *animals* datasets.

---
[4]Traditionally, $\kappa$ calculates priors for each label to compensate for

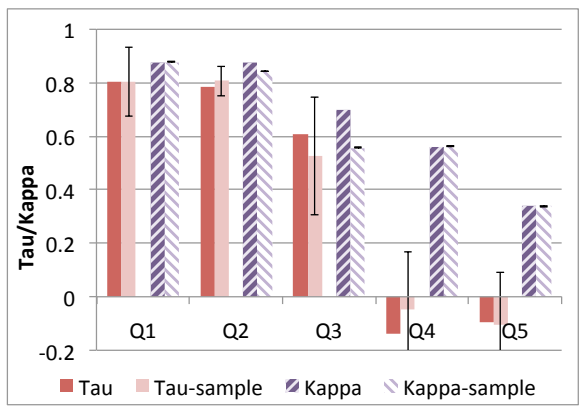

Figure 6: $\tau$ and $\kappa$ metrics on 5 queries.

We generated five queries that represent five sort tasks:
**Q1**: Sort squares by size
**Q2**: Sort animals by "Adult size"
**Q3**: Sort animals by "Dangerousness"
**Q4**: Sort animals by "How much this animal belongs on Saturn"
**Q5**: (Artificially) generate random Compare and Rate responses.

The instructions for Q3 and 4 were deliberately left open-ended to increase the ambiguity. Q4 was intended to be a nonsensical query that we hoped would generate random answers. As we describe below, the worker agreement for Q4's *Compare* tasks was higher than Q5, which suggests that even for nonsensical questions, workers will apply and agree on some preconceived sort order.

For lack of objective measures, we use the Compare results as ground truth. The results of running Compare on queries 2, 3, and 4 are as follows:

**Size**: ant, bee, flower, grasshopper, parrot, rock, rat, octopus, skunk, tazmanian devil, turkey, eagle, lemur, hyena, dog, komodo dragon, baboon, wolf, panther, dolphin, elephant seal, moose, tiger, camel, great white shark, hippo, whale
**Dangerousness**: flower, ant, grasshopper, rock, bee, turkey, dolphin, parrot, baboon, rat, tazmanian devil, lemur, camel, octopus, dog, eagle, elephant seal, skunk, hippo, hyena, great white shark, moose, komodo dragon, wolf, tiger, whale, panther
**Belongs on Saturn**[5]: whale, octopus, dolphin, elephant seal, great white shark, bee, flower, grasshopper, hippo, dog, lempur, wolf, moose, camel, hyena, skunk, tazmanian devil, tiger, baboon, eagle, parrot, turkey, rat, panther, komodo dragon, ant, rock

Figure 6 show $\tau$ and $\kappa$ for each of the five queries. Here $\kappa$ is computed on the comparisons produced by the Compare tasks. The figure also shows the effect of computing these metrics on a random sample of 10 of the squares/animals rather than the entire data set (the sample bars are from 50 different samples; error bars show the standard deviation of each metric on these 50 samples.)

The results show that the ambiguous queries have progressively less worker agreement ($\kappa$) and progressively less agreement between sorting and rating ($\tau$). While $\kappa$ decreases between Q3 and Q4 (dangerousness and Saturn), it is not as low in Q4 as it is in Q5 (Saturn and random). While there is little agreement between

---

bias in the dataset (e.g., if there are far more small animals than big animals). We found this doesn't work well for sort-based comparator data due to correlation between comparator values, and so we removed the compensating factor (the denominator in Fleiss' $\kappa$).
[5]Note that while size and dangerousness have relatively stable orders, the Saturn list varies drastically as indicated by low $\kappa$. For example, in three runs of the query, rock appeared at the end, near the beginning, and in the middle of the list.

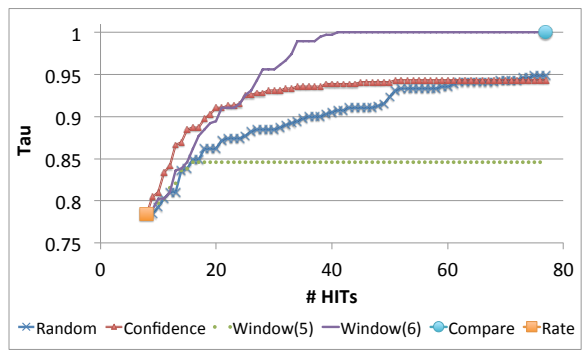

Figure 7: Hybrid sort algorithms on the 40-square dataset.

workers on animals that belong on Saturn, their level of agreement is better than random. For example, Komodo Dragon was consistently rated as highly adaptable to Saturn's environment. The decrease in $\kappa$ with increasing ambiguity suggests that $\kappa$ is a useful signal in identifying unsortable datasets.

$\tau$ is significantly lower for Q4 than for Q3, which suggests that ordering by rating does not work well for Q4 and that we should probably use the Compare method for this workload rather than the Rate method. For Q1, Q2, and Q3, however, Rate agrees reasonably well with Compare, and because it is significantly cheaper, may be a better choice.

Finally, we note that sampling 10 elements is an effective way to estimate both of these metrics, which means that we can run both Rate and Compare on samples, compute $\tau$ and $\kappa$, and decide whether to order the rest of the data set with Rate or Compare (depending on $\tau$), or to stop ordering at that point (if $\kappa$ is very low.)

### 4.2.4 Hybrid Approach

Our final set of experiments measure how the hybrid approaches perform in terms of accuracy with increasing number of HITs. We aim to understand how the sort order of hybrid changes between Rank quality and Compare quality with each additional HIT.

The first experiment uses the 40-square dataset. The comparison interface shows 5 items at a time. We set window size $S = 5$ to be consistent with the number of items in a single comparison HIT. Figure 7 shows how $\tau$ improves with each additional HIT. Compare (upper right circle) orders the list perfectly, but costs 78 HITs to complete. In contrast, Rate (lower left square) achieves $\tau = 0.78$, but only costs 8 HITs (batch size=5). In addition to these two extremes, we compared four schemes, based on those described in Section 4.1.3: random, confidence-based, windowing with $t = 5$ (Window 5), and windowing with $t = 6$ (Window 6).

Overall, Window 6 performs best, achieving $\tau > .95$ in under 30 additional HITs, and converging to $\tau = 1$ in half the HITs that Compare requires. Window 5 performs poorly because $t$ is a multiple of the number of squares, so multiple passes over the data set (beyond the 8th HIT) do not improve the ordering. As the list becomes more ordered, random is more likely to compare items that are correctly ordered, and thus wastes comparisons. Confidence does not perform as well as Window 6—prioritizing high-variance regions assists with fixing local sort order mistakes, but does not systematically move items that are far from their correct positions. In the sliding window scheme, after several passes through the dataset items that were far away from their correct position can migrate closer to the correct location.

Finally, we executed Q2 (animal size query) using the hybrid scheme and found similar results between the approaches. Ultimately, the window-based approach performed the best and improved $\tau$ from .76 to .90 within 20 iterations.

| Operator | Optimization | # HITs |
|---|---|---|
| Join | Filter | 43 |
| Join | Filter + Simple | 628 |
| Join | Filter + Naive | 160 |
| Join | Filter + Smart 3x3 | 108 |
| Join | Filter + Smart 5x5 | 66 |
| Join | No Filter + Simple | 1055 |
| Join | No Filter + Naive | 211 |
| Join | No Filter + Smart 5x5 | 43 |
| Order By | Compare | 61 |
| Order By | Rate | 11 |
| Total (unoptimized) | | 1055 + 61 = 1116 |
| Total (optimized) | | 66 + 11 = 77 |

Table 5: Number of HITs for each operator optimization.

## 4.3 Summary

In summary, we presented two sort interfaces and algorithms based on ratings (linear complexity) and comparisons (quadratic complexity). We found that batching is an effective way to reduce the complexity of sort tasks in both interfaces. We found that while significantly cheaper, ratings achieve sort orders close to but not as good as comparisons. Using two metrics, $\tau$ and a modified $\kappa$, we were able to determine when a sort was too ambiguous ($\kappa$) and when rating performs commensurate with comparison ($\tau$).

Using a hybrid window-based approach that started with ratings and refined with comparisons, we were able to get similar ($\tau > .95$) accuracy to sorts at less than one-third the cost.

## 5. END TO END QUERY

In the previous sections, we examined different operator optimizations in isolation. We now execute a complex query that utilizes joins and sorts, and show that our optimizations reduce the number of HITs by a factor of $14.5\times$ compared to a naive approach.

### 5.1 Experimental Setup

The query joins a table of movie frames and a table of actor photos, looking for frames containing only the actor. For each actor, the query finds frames where the actor is the main focus of the frame and orders the frames by how flattering they are:

```
SELECT name, scene.img
FROM actors JOIN scenes
  ON inScene(actors.img, scenes.img)
  AND POSSIBLY numInScene(scenes.img) > 1
ORDER BY name, quality(scenes.img)
```

The query uses three crowd-based UDFs:
**numInScene**, a generative UDF that asks workers to select the number of people in the scene given the options (0, 1, 2, 3+, UN-KNOWN). This UDF was designed to reduce the number of images input into the join operator.
**inScene**, an EquiJoin UDF that shows workers images of actors and scenes and asks the worker to identify pairs of images where the actor is the main focus of the scene.
**quality**, a Rank UDF that asks the worker to sort scene images by how flattering the scenes are. This task is highly subjective.

We tried several variants of each operator. For the numInScene filter we executed feature extraction with batch size 4. We also tried disabling the operator and allowing all scenes to be input to the join operator. For the inScene join, we use Simple, Naive batch 5, and Smart batch 3×3 and 5×5. For the quality sort, we used Compare with group size 5, and Rate batch 5.

For the dataset, we extracted 211 stills one second apart from a three-minute movie. Actor profile photos came from the Web.

### 5.2 Results

The results are summarized in Table 5. The bottom two lines show that a simple approach based on a naive, unfiltered join plus comparisons requires 1116 hits, whereas applying our optimizations reduces this to 77 hits. We make a few observations:
**Smart Join:** Surprisingly, we found that workers were willing to complete a 5x5 SmartJoin, despite its relative complexity. This may suggest that SmartJoin is preferred to naive batching.
**Feature Extraction:** We found that the benefit of numInScene feature extraction was outweighed by its cost, as the the selectivity of the predicate was only 55%, and the total number of HITs required to perform Smart Join with a 5x5 grid was relatively small. This illustrates the need for online selectivity estimation to determine when a crowd-based predicate will be useful.
**Query Accuracy:** The numInScene task was very accurate, resulting in no errors compared to a manually-evaluated filter. The inScene join did less well, as some actors look similar, and some scenes showed actors from the side; we had a small number of false positives, but these were consistent across implementations. Finally, the scene quality operator had high variance and was quite subjective; in such cases *Rate* works just as well as *Compare*.

## 6. DISCUSSION AND FUTURE WORK

In this section, we discuss issues and lessons learned from our implementation and efforts running jobs on Mechanical Turk.
**Reputation:** While not directly related to database implementation, it is important to remember that your identity carries reputation on MTurk. Turkers keep track of good and bad requesters, and share this information on message boards such as Turker Nation[6]. By quickly approving completed work and responding to Turker requests when they contact you with questions, you can generate a good working relationship with Turkers.

When we started as requesters, Turkers asked on Turker Nation if others knew whether we were trustworthy. A Turker responded:

[requester name] is okay .... I don't think you need to worry. He is great on communication, responds to messages and makes changes to the Hits as per our feedback.

Turker feedback is also a signal for price appropriateness. For example, if a requester overpays for work, Turkers will send messages asking for exclusive access to their tasks.
**Choosing Batch Size:** We showed that batching can dramatically reduce the cost of sorts and joins. In studying different batch sizes, we found batch sizes at which workers refused to perform tasks, leaving our assignments uncompleted for hours at a time. As future work, it would be interesting to compare adaptive algorithms for estimating the ideal batch size. Briefly, such an algorithm performs a binary search on the batch size, reducing the size when workers refuse to do work or accuracy drops, and increasing the size when no noticeable change to latency and accuracy is observed.

As a word of caution, the process of adaptively finding the appropriate batch sizes can lead to worker frustration. The same Turker that initially praised us in Section 6 became frustrated enough to list us on Turker Nation's "Hall of Shame:"

These are the "Compare celebrity pictures" Hits where you had to compare two pictures and say whether they were of the same person. The Hit paid a cent each. Now there are 5 pairs of pictures to be checked for the same pay. Another Requester reducing the pay drastically.

---
[6] http://turkers.proboards.com/

Hence, batching has to be applied carefully. Over time, ideal starting batch sizes can be learned for various media types, such as joins on images vs. joins on videos.

**Worker Selection:** We found that the QA method of Ipeirotis et al. [6] works better than simple majority vote for combining multiple assignment answers and is able to effectively eliminate and identify workers who generate spam answers. Majority vote can be badly skewed by low-quality answers and spam.

To allow us to compare across experiments, we elected not to ban workers from completing future tasks even if they were clearly generating poor output. In a non-experimental scenario, one could use the output of the QA algorithm to ban Turkers found to produce poor results, reducing future costs.

One limitation of QA is that it is designed for categorical data, when workers assign categories to records. Devising a similar method for ordinal and interval data is interesting future work.

**Scaling Up Datasets:** In our experiments, we described techniques that provide order-of-magnitude cost reductions in executing joins and sorts. Still, scaling datasets by another order of magnitude or two would result in prohibitive costs due to the quadratic complexity of both join and sort tasks. Hence, one important area of future work is to integrate human computation and machine learning, training classifiers to perform some of the work, and leaving humans to peform the more difficult tasks.

**Whole Plan Budget Allocation:** We have described how Qurk can determine and optimize the costs of individual query operators. Another important problem is how to assign a fixed amount of money to an entire query plan. Additionally, when there is too much data to process given a budget, we would like Qurk to be able to decide which data items to process in more detail.

**Iterative Debugging:** In implementing queries in Qurk, we found that workflows frequently failed because of poor Turker interface design or the wording of a question. Crowd-powered workflow engines could benefit from tools for iterative debugging.

## 7. RELATED WORK

The database community has recently seen increasing interest in crowdsourced databases. Franklin et al. described CrowdDB [5], a database with a declarative interface and operators for handling joins, sorts, and generative queries in their data definition language. Their experiments explore the properties of MTurk and show the feasibility of joins and sorts, but they do not provide a detailed discussion of implementation alternatives or performance tradeoffs. Our contribution is to study how to achieve order-of-magnitude price improvements while maintaining result accuracy. Parameswaran et al. [13] also proposed a crowdsourced database including a Datalog-based query language for querying humans, and provide some thoughts on how to reason about uncertain worker responses.

Systems for posting tasks to MTurk are available outside the databases community. TurKit [10] is a system and programming model by Little et al. that allows developers to iteratively build MTurk-based applications while caching HIT results between program runs. Kittur et al.e [8] described CrowdForge, a MapReduce-style model for large task decomposition and verification.

Because we retrieve multiple worker responses to each question, we must decide how to arrive at the correct answer given several. A simple approach, used by CrowdFlower[7], is to require gold standard data with which to test worker quality, and ban workers who perform poorly on the gold standard. For categorical data, selecting a majority vote of responses is also powerful. Dawid and Skene presented an expectation maximization technique for for iteratively estimating worker and result quality [3] in the absense of gold standard data. Ipeirotis et al. [6] modified this technique to consider bias between workers in addition to worker quality on categorical data. We utilize this technique to improve join results.

Mason and Watts [12] studied the effects of price on quantity and quality of work. They find that workers are willing to perform more tasks when paid more. They also find that for a given task difficulty, result accuracy is not improved by increasing worker wages. This leads to our experiment design choice of studying how to reduce the number of HITs while maintaining accuracy and per-HIT cost.

## 8. CONCLUSION

We presented an approach for executing joins and sorts in a declarative database where humans are employed to process records. Our system, Qurk, runs on top of crowdsourcing platforms like MTurk. For join comparisons, we developed three different UIs (simple, naive batching, and smart batching), and showed that the batching interfaces can reduce the total number of HITs to compute a join by an order of magnitude. We showed that feature filtering can pre-filter join results to avoid cross products, and that our system can automatically select the best features. We presented three approaches to sorting: comparison-based, rating-based, and a hybrid of the two. We showed that rating often does comparably to pairwise comparisons, using far fewer HITs, and presented signals $\kappa$ and $\tau$ that can be used to determine if a data set is sortable, and how well rating performs relative to comparison. We also present a hybrid scheme that uses a combination of rating and comparing to produce a more accurate result than rating while using fewer HITs than comparison. We showed on several real-world datasets that we can greatly reduce the total cost of queries without sacrificing accuracy – we reduced the cost of a join on a celebrity image data set from $67 to about $3, a sort by $2\times$ the worst-case cost, and reduced the cost of an end-to-end example by a factor of 14.5.

---
[7]http://crowdflower.com